\documentclass[aps,pra,10pt,twocolumn,superscriptaddress]{revtex4-2}

\usepackage[dvipsnames]{xcolor}
\usepackage{hyperref}
\hypersetup{
  breaklinks=true,
  colorlinks=true,
  allcolors=BlueViolet,
}

\usepackage{graphicx} % inserting figures
\graphicspath{{./figures/}} % set path for all figures

\usepackage[inline]{enumitem}  % inline enumeration
\setlist{nosep} % more compact spacing between environments

\usepackage{physics}  % basic math
\usepackage{qcircuit}
\usepackage{xy} % for xy-pic diagrams
\usepackage{bm}  % bold math with \bm
\usepackage{braket}  % braket notation
\usepackage{amssymb}  % for the \intercal / transpose symbol
\usepackage{amsmath}
\usepackage{mathdots}  % for upward-going diagonal dots: \iddots
\usepackage{mathtools}  % or \coloneqq
\usepackage{float}
\renewcommand{\arraystretch}{0.6}

\usepackage{geometry}    
\makeatletter

\begin{document}

\title{
Quantum Depth Compression via Local Dynamic Circuits}
\author{Benjamin Hall}
\affiliation{Infleqtion (Chicago, IL, USA)}
% \email{ben.hall@infleqtion.com}
\author{Palash Goiporia}
\affiliation{Infleqtion (Chicago, IL, USA)}
\author{Rich Rines}
\affiliation{Infleqtion (Chicago, IL, USA)}

\begin{abstract}

We present Quantum Depth Compression (QDC), a general compilation framework that utilizes dynamic circuits to reduce arbitrary quantum circuits to depth linear in the number of non-Clifford gates and to grid connectivity without the need for expensive SWAP-networks. The framework consists of pushing Clifford gates to the end of the circuit, resulting in a sequence of non-Clifford Pauli-phasors followed by an all Clifford sub-circuit, both of which are then reduced to constant depth via dynamic circuits. We show that applying QDC to random Pauli-phasor circuits lowers both their depth and CNOT count compared to a standard alternative compiler. %Finally, we present noisy simulations showing that compiling the same circuits via QDC also increases their fidelity compared to the same alternative.
\end{abstract}

\maketitle

\section{Introduction}

Quantum computers are fundamentally limited by space and time.
The Hilbert space explorable by the computer is determined by its number of qubits, while the time it takes to execute a computation is set by its native gate speeds. 
Some quantum computing modalities are more space-limited than time-limited (e.g. superconducting and spin-qubit based quantum computers). 
For example, while superconducting devices have fast gate times (10 - 100ns \cite{ibm_gate_speed}), they struggle to scale their qubit count due to engineering constraints such as control wiring, cross-talk, and cryogenic cooling requirements. 
Other modalities are more time-limited than space-limited (e.g. neutral atom and trapped ion based quantum computers). For example, while neutral atom devices are scalable to large numbers of qubits (with a current record of 6,100 \cite{neutral_atom_qubit_record}), they have long gate times (0.1 - 1$\mu$s \cite{neutral_atom_gate_speed}). 

Quantum circuits, which program quantum computers, are also describable via space and time.
The width of the circuit (number of qubits) determines the size of its explorable Hilbert space, while the depth of the circuit (number of sequential gates layers) determines the time it will take to be executed. A quantum circuit's space and time (width and depth) requirements are dictated by the problem it addresses. Thus, quantum computing modalities (which have fixed space-time constraints) are often limited in the applications they can execute efficiently by the space-time dimensions of the required circuit.

However, there exists in quantum computing, a space-time tradeoff, which may be exploited to compress the depth of a circuit (time) at the cost of expanding its width (space). Such a tradeoff is particularly enticing for time-limited quantum modalities like neutral atoms and trapped-ions. This space-time tradeoff is mediated through entanglement and projection, which allow for the parallelization of the Clifford portions of quantum circuits. While entanglement is straightforward in standard circuits, performing true projection is impossible and pseudo-projection via post-selection (as required with standard circuits) is exponentially costly. However, dynamic circuits, which admit mid-circuit measurement and classical feed-forward operations, enable corrections of undesired measurements, thus enabling pseudo-projection without said exponential overhead. These corrections are able to be done after measurement via the teleportation enabled by entanglement.

Previous techniques also exploit this space-time tradeoff, such as measurement-based quantum computation \cite{mbqc} and gate teleportation \cite{gate_teleportation}. While enticing, these techniques require large entangled resource states and, often, highly non-local connectivity. Dynamic circuits exist as a hybrid between standard and measurement-based quantum computation. They are computationally similar to standard circuits, but use measurement information to inform part of their computation.

Regardless of modality, short depth is vital for both near-term intermediate scale quantum (NISQ) and fault-tolerant applications. Most NISQ era algorithms are variational quantum algorithms, which exhibit a trainability phase transition: circuits whose depth scales extensively with system size have exponentially vanishing gradients \cite{mcclean2018barren}, while circuits with limited depth remain optimizable \cite{holmes2022expressibility}. For fault-tolerant circuits, compressing depth reduces both the number of error-correction cycles, lowering logical error accumulation, and the program runtime, which is crucial for achieving speed-up over classical computation.

Routing quantum circuits to match device connectivity is also costly in terms of both time and space. If physical qubits are immobile (e.g. superconducting) then routing must be done via costly SWAP-networks which increase both depth and two-qubit entangling gate count, both major sources of noise. If physical qubits are mobile (e.g. neutral atoms) then routing can be done through movement which, while more flexible, not only increases circuit runtime but also introduces errors of its own. Thus, regardless of modality, efficient routing is crucial.

Here, we present Quantum Depth Compression (QDC), a compilation algorithm that exploits this space-time tradeoff to compress the depth of an arbitrary quantum circuit to depth linear in the number of non-Clifford gates via dynamic circuits. The compilation also results in a circuit requiring only 2D-local (grid) connectivity. It achieves this by combining, synthesizing, and reducing the Clifford sections of the circuit to constant-depth via dynamic circuits. Thus, QDC's depth compressibility increases with the Cliffordness of the circuit and is therefore particularly useful for highly Clifford physical circuits (e.g. VQA's with Clifford parameter snapping \cite{cafqa}) and for most logical circuits (as many quantum error correcting codes are dominated by Clifford gates \cite{gottesman1997stabilizer}). 

\section{Overview}

\begin{figure}[t]
    \centering
    \includegraphics[width=1\linewidth]{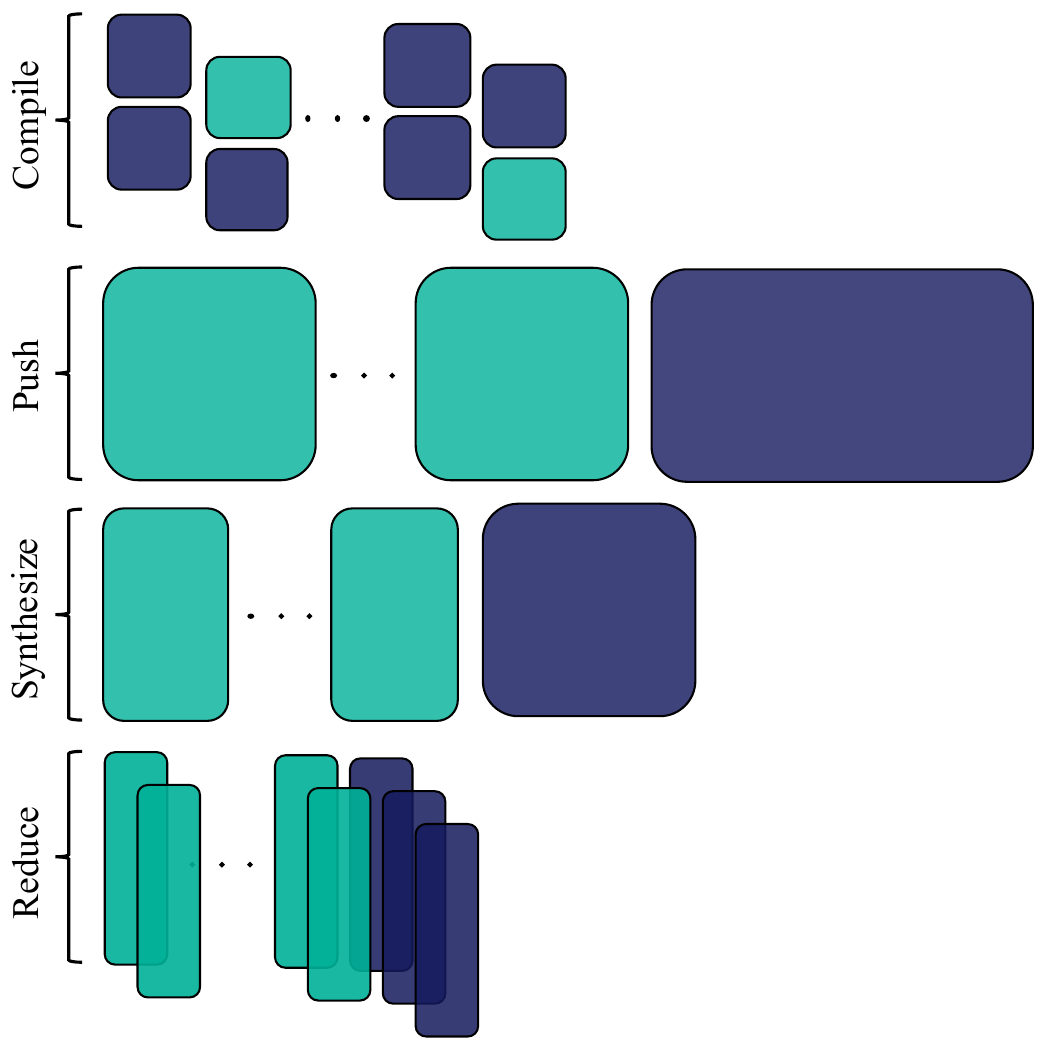}
    \caption{Overview of QDC's steps. Compile: each blue and turquoise square represents a Clifford section $C_i$ and non-Clifford Pauli-phasor $P_i$, respectively. Push: each turquoise square and blue rectangle represents a modified Pauli-phasor $P'_i$ or combined Clifford section $C$, respectively. Synthesize: the blue square represents the synthesized Clifford section $C'$. Reduce: see Figure \ref{fig:reduce} below.}
    \label{fig:fig1}
\end{figure}

The QDC algorithm consists of the following steps, as illustrated in Figure \ref{fig:fig1} on the right:
\\
\begin{enumerate}
\item \textbf{Compile} the circuit into Clifford sections $C_i$ and non-Clifford Pauli-phasors $P_i$.
\begin{align*}
\prod_{i}P_iC_i.
\end{align*}
\item \textbf{Push} all Clifford sections $C_i$ to the end of the circuit, modifying each Pauli-phasor $P_i\to P'_i$ in the process
\begin{align*}
C\prod_jP'_i
\end{align*}
where $C=\prod_iC_i$.
\item \textbf{Synthesize} the combined Clifford section $C\to C'$ to 2D-local connectivity and depth linear in the number of qubits.
\begin{align*}
C'\prod_iP'_i.
\end{align*}
\item \textbf{Reduce} both the decomposed Clifford section $C'\to C''$ and each modified Pauli-phasor $P'_i\to P''_i$ to constant depth via local dynamic circuits.
\begin{align*}
C''\prod_iP''_i
\end{align*}
\end{enumerate}

\begin{figure}[H]
    \centering
    \includegraphics[width=1\linewidth]{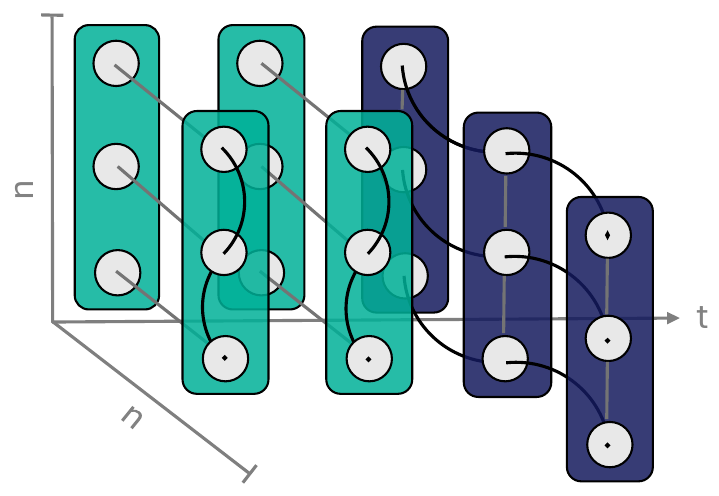}
    \caption{An $n=3$ qubit example of the fourth QDC step (reduce) showing steps in time (t) of an $\text{n}\times\text{n}$ grid of qubits (gray circles). 
    The turquoise and blue time steps represent reduced Pauli-phasor and Clifford sections, respectively. The black and gray curves represent cups/caps (Figure \ref{fig:cup_and_cap}) and two-qubit entangling gates, respectively.}
    \label{fig:reduce}
\end{figure}

Figure \ref{fig:reduce} provides a ``zoomed in'' version of the Reduce section of Figure \ref{fig:fig1}. The black and gray lines represent cups/caps (Figure \ref{fig:cup_and_cap}) and the two-qubit entangling gates of the original circuit, respectively. In both figures, and throughout the paper, blue and turquoise are used to represent Clifford subcircuits and (non-Clifford) Pauli-phasors, respectively.

The next four sections (Section \ref{sec:compile} - \ref{sec:reduce}) walk through each of these steps in detail, highlighting the important role each step plays in the overall algorithm. The reduce section (Section \ref{sec:reduce}) in particular, presents novel reductions of both Pauli-phasors and Clifford circuits to constant depth via 2D-local dynamic circuits that improve over previous work (Table \ref{tab:pervious_work}) in terms of both depth and CX count.
Then, in Section \ref{sec:application}, the QDC is applied to an example circuit consisting of random Pauli-phasors, which are ubiquitous in many algorithms (e.g. QAOA \cite{qaoa}, VQE \cite{vqe}, QML \cite{qml}, and time-evolution). 
As the extent of the depth compression achievable by QDC depends on the Cliffordness of the circuit, we specifically consider the case in which we can ``snap'' a large fraction of the Pauli-phasors to be Clifford (e.g. CAFQA \cite{cafqa}). Finally, we present numerics (Section \ref{sec:numerics}) showing how both the final depth and CNOT count are reduced via QDC compared to a standard decomposition.
%Subsection \ref{sec:numerics} 
%Finally, Subsection \ref{sec:noisy}) presents noisy simulations showing increased fidelity of the same example circuits when compiled via QDC versus the same standard decomposition.

Altogether, QDC presents an efficient way to compress the depth of arbitrary quantum circuits in a manner that only requires limited connectivity. Though the algorithm is particularly useful for time-limited quantum hardware and highly Clifford quantum circuits, Section \ref{sec:numerics} shows that QDC reduces depth without significantly increasing CX count for all Clifford levels. Along with the novel reduction methods presented for both Pauli-phasors and Clifford subcircuits, QDC's depth compression via grid connectivity requirements represents a novel contribution to quantum circuit compilation.

\section{Compile}
\label{sec:compile}

\begin{figure}[H]
    \centering
    \includegraphics[width=1\linewidth]{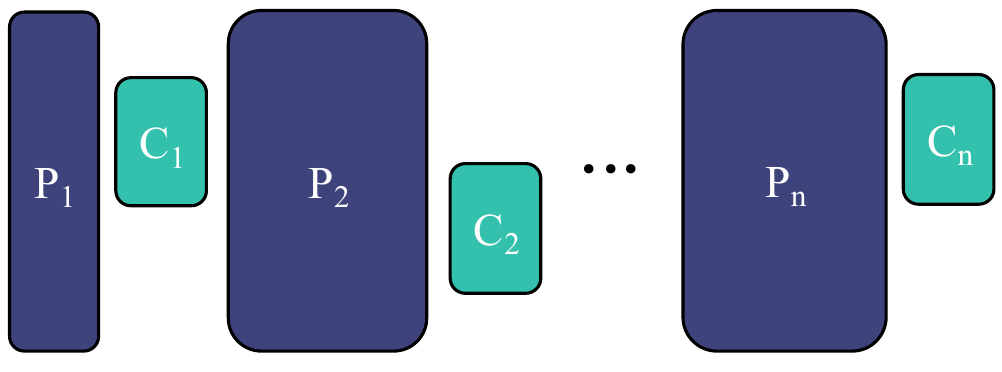}
    \caption{Depiction of the circuit after the compile step. Blue and turquoise rectangles represent Clifford sections and non-Clifford Pauli-phasors, respectively.}
    \label{fig:compile}
\end{figure}

The first step of QDC is to compile the input quantum circuit into Clifford sections $C_i$ and non-Clifford Pauli-phasors $P_i$ (as shown in Figure \ref{fig:compile}):
\begin{align}
\prod_{i}P_iC_i.
\end{align}
Any quantum circuit can be decomposed into a universal gate set, such as $\{R_x, R_z, \text{CNOT}\}$. The single qubit rotations $R_x(\theta)$ and $R_z(\theta)$ are either Clifford, if $\theta=n\pi/2$ for $n\in\mathcal{N}$, or non-Clifford, otherwise. Thus, the CNOTs and Clifford single qubit rotations form Clifford sections $C_i$ while the other single qubit rotations are non-Clifford Pauli-phasors $P_i$, as summarized below: 
\begin{align*}
&\text{Clifford sections ($C$):} \\
&\{R_x(\theta), \ R_z(\theta) : \theta=n\pi/2 \ \forall \ n \in \mathbb{N}\}
+\{\text{CNOT}\}
\\
\\
&\text{Non-Clifford Pauli-phasors ($P$):} \\
&\{R_x(\theta), \ R_z(\theta) : \theta\neq n\pi/2 \ \forall \ n \in \mathbb{N}\}
\end{align*}

Each individual non-Clifford rotation is counted as one non-Clifford Pauli-phasor. Clifford sections consist of all Clifford gates in between non-Clifford rotation gates. In a moment (collection of gates executed in parallel) consisting of both Clifford gates and non-Clifford Pauli-phasors, the Clifford gates are considered without loss of generality, to be part of the Clifford section after the Pauli-phasors in the same moment, as all Cliffords will eventually be pushed to the end of the circuit. Note that some non-Clifford Pauli-phasors have no Clifford section between them, which doesn't affect the next step, push.

Note that if the input circuit is already a sequence of Pauli-phasors, as will be considered in Section \ref{sec:application} (Application), then the compilation step may be skipped entirely as each Pauli-phasor is either Clifford (a Clifford section) or non-Clifford (a non-Clifford Pauli-phasor). Finally, note that if the input circuit is already given in terms of a universal gate set of the following form:
\begin{align*}
&\text{Universal gate-set:} \\
&\{R_\text{p}(\theta), \ R_q(\theta), \ \text{any Clifford 2-qubit entangling gate}\}
\end{align*}
where $p\neq q \in \{X, Y, Z\}$, then the compile step may also be skipped and treated analogously to the $\{R_x, R_z, \text{CNOT}\}$ case.

\section{Push}

\begin{figure}[H]
    \centering
    \includegraphics[width=1\linewidth]{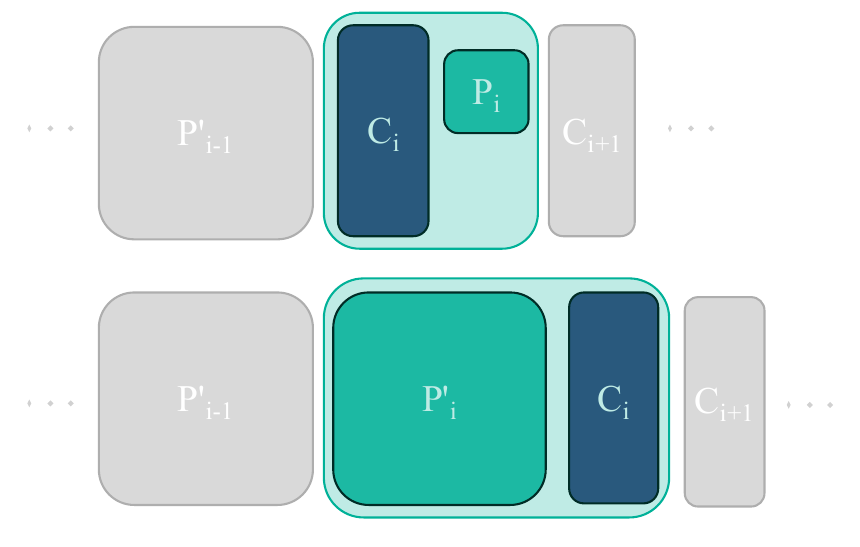}
    \caption{Example of one iteration of the push step: the left-most Clifford section $C_i$ is pushed past the Pauli-phasor $P_i$ immediately to its right, modifying the Pauli-phasor to $P'$ in the process. }
    \label{fig:push}
\end{figure}

The second step of QDC is to push each Clifford section ($C_i$) past its postceding Pauli-phasors ($P_j$ for $j\geq i$), as depicted in Figure \ref{fig:push}, resulting in all the Clifford sections being collected at end of the circuit, preceded by conjugated Pauli-phasors $P'_j$ (Figure \ref{fig:fig1}b):
\begin{align*}
\prod_{i=n}^1 P_iC_i \to C\prod_{j=n}^1P'_j
\end{align*}
where $\prod_{i=n}^1C_i$ and 
$P'_j = \left(\prod_{k=1}^jC_k^\dagger\right) P_j \left(\prod_{k=j}^1C_j\right)$, as depicted in the Push section of Figure \ref{fig:fig1}.

By definition, a Clifford operation $C$ may be ``pushed'' past a Pauli-string $p=\otimes_ip_i$, where $p_i\in\{I, X, Y, Z\}$, as follows: $Cp = p'C$, modifying the Pauli-string from $p\to p'=C p C^\dagger$ in the process. The same holds for Pauli-phasors $P$: $CP=P'C$, as they are exponentiated Pauli-strings: $P=\exp\{i\theta p\}$.

Note that the modified Pauli-phasors $P_i'$ in Figure \ref{fig:push} are wider than their unmodified counter parts $P_i$. This is because the unmodified Pauli-phasors $P_i$ are weight-one Pauli-phasors (exponentiations of Pauli-strings with exactly one Pauli) whereas $P_i'$ may be of larger weight after conjugation, therefore requiring a longer decomposition. This increase in depth is undone during the reduction step (Section \ref{sec:reduce}) which reduces each Pauli-phasor to constant depth. 

One may inquire as to why one should bother with applying QDC at all if there are no depth savings for the non-Clifford Pauli-phasors. First, applying QDC collects all Clifford gates and reduces them to constant depth (Section \ref{sec:reduce}), resulting in a circuit linear in just the number of non-Clifford gates rather than Clifford sections. Second, if the circuit is already in terms of Pauli-phasors, as is common in many circuits (Section \ref{sec:application}) then the modified Pauli-phasors $P'_i$ are not necessarily higher weight than their unmodified counterparts $P_i$. And third, QDC also serves as a constant-depth router to grid connectivity. These savings are borne out in the numerics of our test case (Section \ref{sec:numerics}).%These savings are borne out in both the numerics and noisy simulations of our test case (Section \ref{sec:application}).

\section{Synthesize}
\label{sec:synthesize}

The third step of QDC is to synthesize the Clifford section $C$ at the end of the circuit to 2D-local connectivity and depth linear in the number of qubits. We first note that the Clifford section $C$ may be excluded entirely from the circuit if one wishes to extract only expectation values of Pauli-strings \cite{quclear}, as is common in many quantum algorithms (e.g. QAOA \cite{qaoa}, VQE \cite{vqe}, and QRBM \cite{qrbm}, etc.) This is because the Clifford section $C$ simply rotates the desired Pauli-string $P$ to a new Pauli-string $P'=C^\dagger P C$ as shown below
\begin{align*}
\bra{\psi}P\ket{\psi}
&=
\bra{\psi'}C^\dagger P C\ket{\psi'} \\
&=
\bra{\psi'}P'\ket{\psi'}
\end{align*}
where $\ket{\psi}=C\ket{\psi'}$ is the state of the quantum circuit after the pushing step and $\ket{\psi'}=\prod_iP'_i\ket{0}$ is the intermediate state immediately before the Clifford section $C$ is applied. Thus, for each Pauli-string $P$ one wishes to estimate the expectation value of, one simply drops $C$ from the circuit and estimates the expectation value of $P'=C^\dagger P C$ instead.

If, on the other hand, one wishes to extract other information from the circuit (such as bit-string counts), then QDC process with the following synthesis:
Any Clifford circuit may be decomposed into a circuit with linear nearest-neighbor (lnn) connectivity and depth linear in the number of qubits \cite{cliff_synth}. One such decomposition decomposes the Clifford circuits $C$ in the the following two subsections: 
\begin{align*}
C = -\ CZ-CX- \ | \ -H-S-CZ-H-P-
\end{align*} where $CZ$, $CX$, $H$, $S$ and $P$ are layers of controlled-$Z$, controlled-$X$, Hadamard, $S$, and phase gates, respectively. The two subsections may be implemented in depth $\sim {5n}$ and $\sim{2n}$, respectively, where $n$ is the number of qubits in the circuit. Noting that the first subsection acts trivially on the all-zero state, QDC pushes it to the beginning of the circuit (potentially changing the Pauli-phasors $P'_i$ into other Pauli-phasors along the way) where after it can be deleted. Thus, only the second subsection must be implemented which only requires depth $\mathcal{O}({2n})$. Finally, QDC also drops the $P$ section at the end of the second subsection as it is at the end of the circuit as phasor gates do not effect bit-string measurements.

Because we have pushed and collected together all Clifford sections of the original circuit, the depth savings of this synthesis are increased over simply synthesizing each Clifford section separately. This is because, without pushing, each Clifford section only has its depth reduced if its depth was already greater than $\sim 2n$. And, even if every Clifford section did have an initial depth greater than $\sim 2n$, the total depth of the Clifford part of the circuit would be linear in the number of Clifford sections, rather than constant. That is:
\begin{align}
\sum_{i=1}^{m}\min\{{d_i, \sim 2n}\} \leq \ \sim 2nm
\end{align}
where $m$ is the number of Clifford sections and $d_i$ are the depths of said Clifford sections.

\section{Reduce}
\label{sec:reduce}

The fourth and final step of QDC is to reduce both the Clifford section and the Pauli-phasors to constant depth via 2D-local dynamic circuits.

\subsection{Clifford Section Reduction}

We now detail how QDC reduces the now linearly-connected  Clifford section $C'$ at the end of the circuit to constant depth using only grid connectivity. Said Clifford circuit ($C$) has $n$ qubits and $m=\mathcal{O}(n)$ depth. Divide $C$ into an odd number $d=\mathcal{O}(n)$ of depth $m/d=\mathcal{O}(1)$ sub-circuits,

% \begin{align*}
% \Qcircuit @C=0.5em @R=1em{\qw & \gate{C} & \qw & & \raisebox{0em}{=} & & & \qw & \gate{C_1} & \gate{C_2} & \qw & & \raisebox{0em}{$\cdots$} & & & \gate{C_{2n+1}} & \qw \\}    
% \end{align*}
\begin{figure}[h!]
    \centering
    \includegraphics[width=0.7\linewidth]{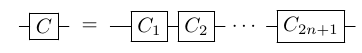}
    \label{fig:placeholder}
\end{figure}

and reduce it to constant depth via ``snaking'':

\begin{figure}[h!]
    \centering
    \includegraphics[width=0.9\linewidth]{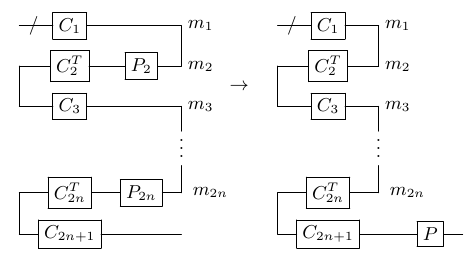}
    \caption{A depth $d$ Clifford circuit $C$ divided in to an odd number of sub-circuits $C_1,...,C_{2n+1}$ executed in constant depth by ``snaking'' the circuit across $nd$ qubits via cups and caps. 
    The required Pauli-corrections $P_2,...,P_{2n}$ may be ``pushed'' to the end of the Circuit resulting in a single combined correct $P$ (Equation \ref{eq:P}).}
    \label{fig:reduce_cliff}
\end{figure}

where $P_{a} = Z^{m_{a-1}}X^{m_{a}}$ are Pauli corrections and the curves are snaking cups/caps, as defined below:

\begin{figure}[h!]
    \centering
    \includegraphics[width=0.9\linewidth]{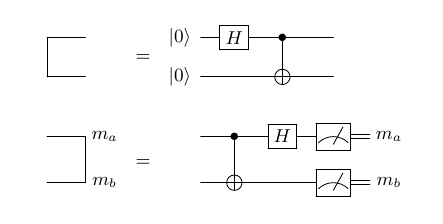}
    \caption{(Top) Cup: Preparation of the Bell-state $\ket{\phi^+}$.
    (Bottom) Cap: Measurement in the same Bell-state basis.}
    \label{fig:cup_and_cap}
\end{figure}

% \begin{figure}[H]
%     \centering
%     \begin{align*}
%     \Qcircuit @C=1em @R=1em{
%     \qwx[1] \qw & \qw & \qw         &     &     &                    & & & \lstick{\ket{0}} & \gate{H} & \ctrl{1} & \qw \\
%             \qw & \qw & \qw         &     &     & \raisebox{2em}{=}  & & & \lstick{\ket{0}} & \qw      & \targ    & \qw \\
%     \\   
%     \qw         & \qw & \qw \qwx[1] & m_a &     &                    & & &                  & \ctrl{1} & \gate{H} & \meter         & \cw & m_a \\
%     \qw         & \qw & \qw         & m_b &     & \raisebox{2em}{=}  & & &                  & \targ    & \qw      & \meter         & \cw & m_b \\
%     }
%     \end{align*}
%     \caption{(Top) Cup: Preparation of the Bell-state $\ket{\phi^+}$.
%     (Bottom) Cap: Measurement in the same Bell-state basis.}
%     \label{fig:cup_and_cap}
% \end{figure}

The Pauli corrections ($P_2$ through $P_{2n}$) of the LHS circuit of Figure \ref{fig:reduce_cliff} are pushed past the Clifford gates ($C_1$ through $C_{2n + 1}$) resulting in a single Pauli correction 
\begin{align}
\label{eq:P}
P=\prod_{i=n}^1\left(C_{2i+1}C_{2i}P_{2i}\right)\prod_{j=1}^{2n+1}C_j^\dagger
.
\end{align}

% Because the original Clifford section $C'$ was of depth $2n+2$ (Section \ref{sec:synthesize}), an $n\times n$ grid of qubits $d=n$ would admit a reduction to constant depth $2 + 2/n$.

\begin{figure*}[t]
    \centering
    \includegraphics[width=\linewidth]{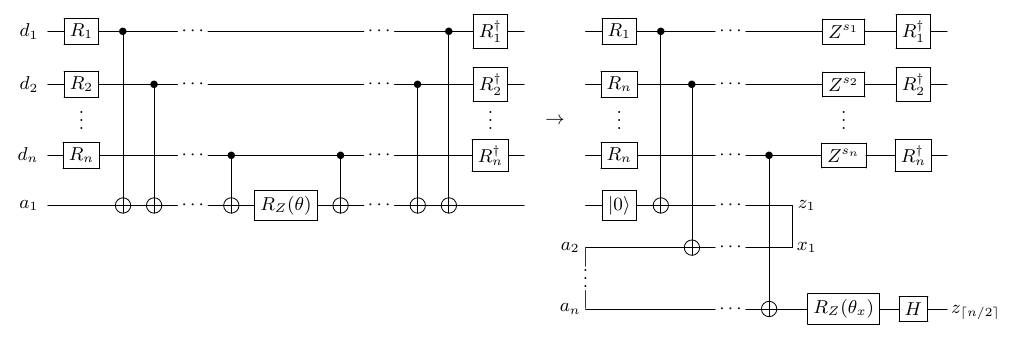}
    \caption{(LHS) A star decomposition of an arbitrary weight-$n$ Pauli-phasor circuit with $n+1$ qubits and depth $\mathcal{O}(n)$. (RHS) QDC's constant-depth circuit decomposition of the same Pauli-phasor with more qubits, $2n$, but only depth $\mathcal{O}(1)$. 
    Here, $\theta_x=(-1)^{\sum_ix_i}\theta$ and $s_j= \sum_{i=\lfloor{j/2}\rfloor}^{\lfloor{n/2}\rfloor}z_{i+1}$.}
    \label{fig:pauli_phasor}
\end{figure*}

\subsection{Pauli-phasor Reduction}

QDC's reduction of a Pauli-phasor $\exp\{i\theta P_1...P_n\}$ starts with its star decomposition (LHS of Figure \ref{fig:pauli_phasor}) which consists of $n$ data qubits $d_1,...,d_n$ and one ancilla qubit $a_1$. Here, a column of basis rotation gates $R_i\in\{I, H, HS^\dagger\}$ is used to rotate from the the $Z\cdots Z$ to $P_1...P_n$ basis, two CNOT starts are used to place the summation of the data qubits on an ancilla qubit, an $R_Z$ gate is used to rotate the state, and then the circuit is mirrored. 

This standard implementation has depth $\mathcal{O}(2n)$ due to the two CNOT stars consisting of $n$ CNOTs being implemented in series. This depth can be reduced to $\mathcal{O}(1)$ by utilizing dynamic circuits to implement the CNOTs of the first star in parallel and eliminating the second CNOT star entirely, RHS of Figure \ref{fig:pauli_phasor}. 
The second CNOT star can be eliminated with dynamics by measuring the ancilla in the $X$ basis and using the principle of deferred measurement to move the measurement before the CNOTs so it occurs immediately after the $R_Z$ gate, applying classically controlled $Z$s on the data qubits $d_1,...,d_n$.
The CNOTs of the first star are parallelized by ``snaking'' the ancilla qubit across $n$ ancilla qubits $a_1,...,a_n$ via cups and caps (Figure \ref{fig:cup_and_cap}). The CNOT targets are then ``ricocheted'' throughout the snake (Appendix \ref{sec:partial_ricochet}).

Without correction, the two circuits of Figure \ref{fig:pauli_phasor} are only equivalent if every cap is measured to be $\ket{00}$. As this is not guaranteed, we introduce a Pauli-correction $P_{z_ax_a}=Z^{z_a}X^{x_a}$ immediately preceding the $a^{\text{th}}$ cap for all caps. These corrections may be pushed down the snake, the $X$s commuting with the targets of the CNOTs and the $Z$s picking up an additional $Z$ on data qubit $d_i$ when pushing past the target of the CNOT between qubits $d_i$ and $a_i$.
The corrections at the end of the snake may then be pushed past the $R_Z$, the $Z$s commuting and the $X$s flipping the sign of the angle $\theta$, allowing us to replace the $\theta$ with $\theta_x=(-1)^{\sum_ix_i}\theta$. After pushing past the $H$, the $X$ corrections (now $Z$ corrections) are eliminated as they now immediately precede a measurement while the $Z$ corrections (now $X$ corrections) are pushed past measurement.

All of these corrections may be combined, resulting in each data qubit $d_j$ simply requiring a correction $Z^{s_j}$ where $s_j = \sum_{i=\lfloor{j/2}\rfloor}^{n-1}z_{i+1}$. These corrections $Z^{s_j}$, being Pauli-phasors themselves, may be pushed rightward within the circuit, past the Clifford $R^\dagger_i$ gates, potentially rotating themselves into other Pauli-phasors. Now at the start of the next reduced Pauli-phasor $P''_{i+1}$, these correction may be pushed past its rotations $R_{i+1}$ as well. This pushing continues until all data qubit corrections are at the end of the last Pauli-phasor, after which they may be pushed through the reduced Clifford section $C''$ to the end of the circuit, where they may be deleted and done in post-processing instead.

When stringing together multiple Pauli-phasors, one would normally perform qubit resets on the ancillas in between reduced Pauli-phasors (LHS of Figure \ref{fig:pauli_phasor}). QDC instead eliminates the need for said resets by replacing them with $X$ corrections that are pushed into the subsequent Pauli-phasor circuit, ultimately being incorporated into its $\theta_x$ and/or $Z^{s_i}$ corrections. The qubit resets of the final Pauli-phasor before the reduced Clifford section $C''$ may similarly be pushed through said reduced Clifford section $C''$ and to the end of the circuit where they may be deleted and done in post-processing instead.

Note that this technique (LHS of Figure \ref{fig:pauli_phasor}) only requires grid connectivity on a $2\times n$ grid of qubits (where the data and ancilla qubits comprise the first and second column, respectively). This is illustrated via the turquoise sections of Figure \ref{fig:reduce}.
If the Pauli-phasor is not ``connected'' (at least one of Paulis $P_2,..,P_{n-1}$ is the identity $I$) then we may simply delete the CNOTs being applied between $d_i$ and $a_i$ if $P_i=I$. We note that implementing a non-connected Pauli-phasor on the same grid via the standard CNOT-ladder method would require expensive SWAP gates which are not required via this technique. 

Note that the Pauli-phasor reduction shown in Figure \ref{fig:pauli_phasor} assumes an odd-weight Pauli-phasor. QDC executes even-weight Pauli-phasors with $n$ ancillas by replacing the bottom cap with a zag, (Figure \ref{fig:zag} of Appendix \ref{appendix:zag}) as shown in Figure $\ref{fig:even_ancilla}$.

\section{Application}
\label{sec:application}

To test QDC we apply it to random Pauli-phasor circuits:
\begin{align*}
\prod_i\left(p^{(i)}_1...p^{(i)}_n\right)^\alpha
\end{align*}
a given percent of which are set to be Clifford, as shown in Figure \ref{fig:vqa} below. Said circuits are ubiquitous across many both variational ansatz and non-variational circuits (e.g. time evolution). In the context of variational ansatze, there is strong evidence \cite{cafqa} that, for certain applications, one may ``snap'' a large percentage of the parameters of a variational ansatz to be Clifford during optimization without sacrificing the its performance.
For each Pauli-phasor, its Pauli-string $p_1...p_n$, is chosen by randomly sampling from the set $\{I, X, Y, Z\}$ with replacement, while its exponent $\alpha$ is drawn from either the uniform distribution $[0,2]$ if non-Clifford or the set $\{0, 1/2, 1, 3/2\}$ if Clifford. 
This is depicted in Figure \ref{fig:vqa} below.
The Pauli-phasors may also be restricted to weight $k$ by randomly selecting $k$ qubit indices, each of which admits a random Pauli (uniformly sampled from the set $\{X, Y, Z\}$) within the Pauli-string.

An example of non-variational circuits consisting of Pauli-phasors is that of time evolution circuits as they simulated exponentiated Hamiltonians, $U(t)=\exp{iHt}$, which can often be written in terms of Paulis
\begin{align*}
H=\sum_{i;\alpha}a_iP^{(\alpha)}_i + \sum_{i,j;\alpha,\beta}b_{i,j}P^{(\alpha)}P^{(\beta)}_j + \cdots
\end{align*}
If a large fraction of the Hamiltonian's coefficients would admit Clifford Pauli-phasors (that is, $a_i, b_{ij}, ... \in \{n/4 \ \text{for} \ n \in \mathcal{N}\}$) as is common in many spin-models (e.g. Heisenberg spin chain \cite{heisenberg}), then QDC would also be particularly well-suited for its compilation.

\begin{figure}[H]
    \centering
    \includegraphics[width=1\linewidth]{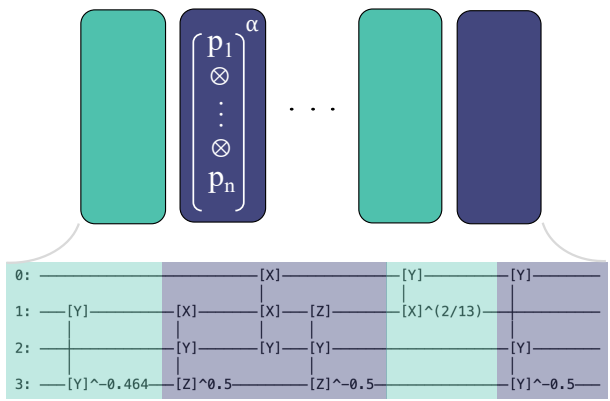}
    \caption{Structure of the circuit used to test QDC, consisting of random Pauli-phasors (Pauli-strings $p_1...p_n$ exponentiated by some exponent $\alpha$), a fixed percentage of which are Clifford (turquoise), the rest of which are non-Clifford (blue). The circuit on the bottom is an example circuit, showing explicit Pauli-phasors, both Clifford and non-Clifford.}
    \label{fig:vqa}
\end{figure}

\pagebreak

\section{Numerics}
\label{sec:numerics}

Here we compute numerical evidence of the utility of QDC by comparing both the depth and CNOT count of our example circuits compiled three different ways, via: standard, QDC (without reduction), and QDC compilation.
Here, standard compilation refers to compiling via Qiskit's \cite{Qiskit} transpilation function with basis gates $\{Rx, \ Rz, \ CX\}$, optimization level 3, and coupling map corresponding to that of a grid.

\begin{figure}[H]
    \centering
    \includegraphics[width=1\linewidth]{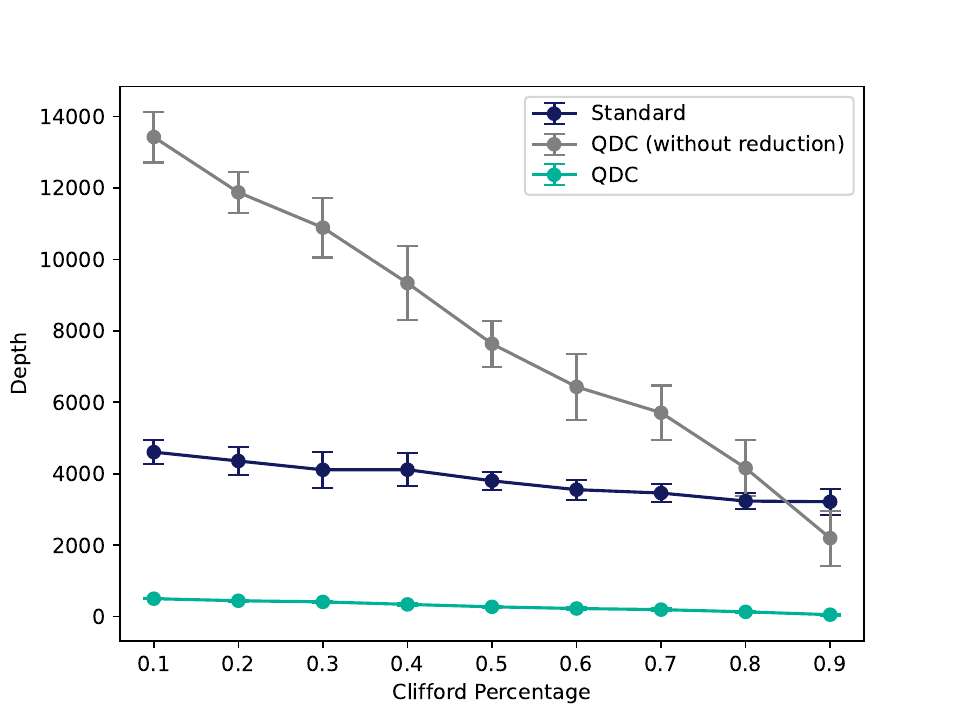}
    \caption{Comparing depth vs Clifford percentage for standard, QDC (without reduction) and QDC compilation.}
    \label{fig:depth_v_cliff}
\end{figure}

\begin{figure}[H]
    \centering
    \includegraphics[width=1\linewidth]{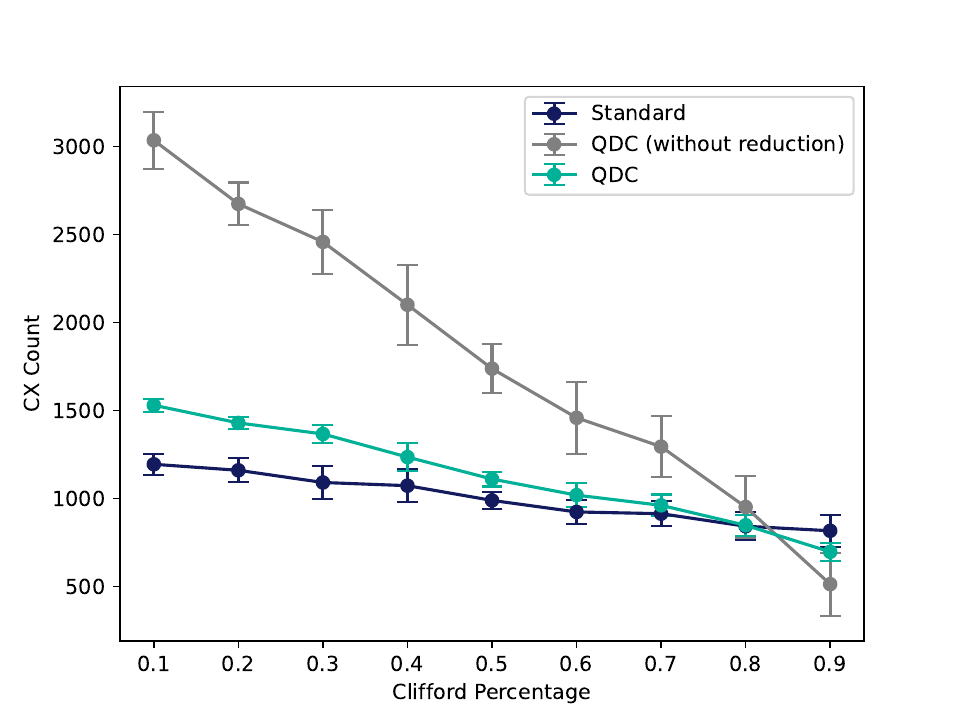}
    \caption{Comparing CX count vs Clifford percentage for standard, QDC (without reduction) and QDC compilation.}
    \label{fig:cx_v_cliff_grid}
\end{figure}

The first numerical experiment was done on 16 qubit circuits (each consisting of 40 random Pauli-phasors) with varying Clifford percentages.
In terms of depth, Figure \ref{fig:depth_v_cliff} shows that while QDC (without reduction) compilation only begins outperforming the standard compilation at 90\% Clifford, full QDC always outperforms standard compilation across all Clifford percentages. In terms of CX counts, Figure \ref{fig:cx_v_cliff_grid} shows that standard compilation does start (statistically significantly) outperforming QDC under 60\% Clifford, but by a much smaller ratio than QDC outperforms standard in terms of depth. 

\begin{figure}[H]
    \centering
    \includegraphics[width=1\linewidth]{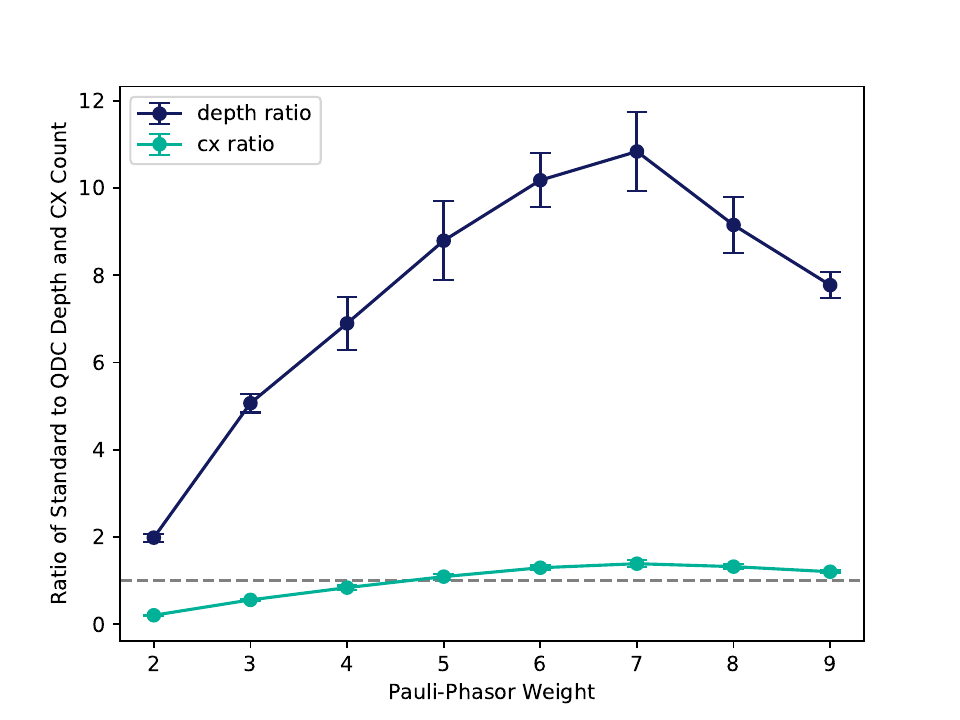}
    \caption{Comparing depth and CX count vs Pauli-phasor weight for standard and QDC compilation, averaged over various Clifford percentages. The gray dashed line indicates a ratio of 1.}
    \label{fig:ratio_v_ppw}
\end{figure}

This is further borne out in the second numerical experiment which were done on 9 qubit circuits (each consisting of 40 random Pauli-phasors) with varying Pauli-phasor weights. In this experiment, we also averaged over the Clifford percentages of the previous experiment. In terms of depth, Figure \ref{fig:ratio_v_ppw} shows that QDC outperforms standard compilation for all Pauli-phasor weights, from a factor of $~$2x for weight 2 to a factor of $~$10x for weight 7. The improvement increases with increasing weight as the depth of linear Pauli-phasor decomposition grows with increased weight. We hypothesize that the improvement begins to diminish after weight as, at these weights, the Pauli-phasors are much less likely to be disconnected and therefore require less routing. In the case of weight 9, in fact, no routing is required as Pauli-phasors of this weight span the entire qubit set. The fact that weight 9 Pauli-phasor circuits still have their depth reduced by QDC over standard decomposition indicates that the depth savings from Pauli-phasor reduction are sufficient to beat standard decomposition and that QDC's efficient routing to grid connectivity simply provides bonus depth savings.

In terms of CX count, we see that QDC only outperforms standard decomposition for Pauli-phasor weights greater than 5. However, we see that the ratio never exceeds 2, which is the minimum depth ratio, showing that QDC reduces depth over standard compilation without a significant change to CX count, which would suggest a lower overall circuit fidelity. We hypothesize that the CX count curve increases before weight 7 and decreases after for a similar reason that the depth curves does, in that the Pauli-phasors increase in their likelihood of being disconnected with increasing depth until it begins to approach the qubit count. Finally, we note that the numerics for this experiment were averaged over various Clifford percentage (those of the first experiment) showing that this analysis holds on average with regard to varying Clifford percentages.

\section{Conclusion}

We have presented Quantum Depth Compression (QDC), a compilation framework that leverages dynamic circuits to reduce the depth of arbitrary quantum circuits while maintaining only 2D-local connectivity. By decomposing circuits into Clifford sections and non-Clifford Pauli-phasors, pushing all Clifford operations to the end, and reducing both types of subcircuits via local dynamic circuits, QDC achieves a depth that scales linearly with the number of non-Clifford gates. 

Our approach is particularly effective for circuits with high Cliffordness, such as variational quantum algorithms with Clifford parameter snapping, as well as for many fault-tolerant logical circuits dominated by Clifford operations. The compilation framework not only reduces circuit depth and CNOT count, but also provides a constant-depth routing solution for grid-connected qubit architectures. In future work, we aim to validate these savings by analyzing noisy simulations of these compiled circuits. %Noisy simulations confirm that these reductions translate to improved circuit fidelity.

Overall, QDC demonstrates that exploiting the space-time tradeoff via dynamic circuits is a practical strategy for optimizing quantum circuits in both NISQ and fault-tolerant regimes. The methods introduced here, especially the reductions of arbitrary Pauli-phasors and Clifford subcircuits, provide a versatile toolkit for efficient quantum compilation, enabling deeper and more reliable computations on hardware with limited connectivity and long gate times.

\section{Acknowledgments}

We thank Michael Perlin for early management of this work. We further thank Stephanie Lee and Teague Tomesh for managing the bulk of this project. We also thank David Owusu-Antwi, Victory Omole, and Teague Tomesh for their feedback on this paper.

This material is based upon work supported by the U.S. Department of Energy, Office of Science, Office of Advanced Scientific Computing Research, under Award Number DE-SC0021526.
 
Disclaimer: This report was prepared as an account of work sponsored by an agency of the United States Government. Neither the United States Government nor any agency thereof, nor any of their employees, makes any warranty, express or implied, or assumes any legal liability or responsibility for the accuracy, completeness, or usefulness of any information, apparatus, product, or process disclosed, or represents that its use would not infringe privately owned rights. Reference herein to any specific commercial product, process, or service by trade name, trademark, manufacturer, or otherwise does not necessarily constitute or imply its endorsement, recommendation, or favoring by the United States Government or any agency thereof. The views and opinions of authors expressed herein do not necessarily state or reflect those of the United States Government or any agency thereof.

\section{Appendix}

\subsection{Comparison to Previous Work}

Table \ref{tab:pervious_work} below compares our work to two previous works on Pauli-phasor and/ore Clifford circuit reduction via dynamic circuits. We have improved the qubit and CX count for Pauli-phasor reduction by a factor of 1.5 and 2, respectively. Additionally, we have improved both the qubit and CX count for Clifford circuit reduction by a factor of 2.

\begin{table}[H]
    \centering
    {\renewcommand{\arraystretch}{1.2}
    \begin{tabular}{cc|c|c|c}
                                        & \multicolumn{2}{c|}{Pauli-phasor}            & \multicolumn{2}{c}{Clifford circuit}         \\
        \cline{2-5}
                                        & \multicolumn{1}{|c|}{\# Qubit}  & \# CX      & \# Qubit   & \multicolumn{1}{c|}{\# CX}       \\
        \hline
        This Work                       & \multicolumn{1}{|c|}{$\sim 2n$} & $\sim n $  & $\sim nm$  & \multicolumn{1}{c|}{$\sim nm$}   \\
        \hline
        Buhrum 2024 \cite{buhrman2024} & \multicolumn{1}{|c|}{$\sim 3n$}  & $\sim 2n$  & $\sim 2nm$ & \multicolumn{1}{c|}{$\sim 2nm$} \\
        \hline
        Yang 2023   \cite{yang2023}    & \multicolumn{1}{|c|}{$\sim 3n$}  & $\sim 2n$  & N/A        & \multicolumn{1}{c|}{N/A       } \\
        \cline{2-5}
    \end{tabular}
    }
    \caption{Comparing this work to previous work (\cite{buhrman2024}, \cite{yang2023}) in terms of qubit and CX count for both Clifford circuit and Pauli-phasor reduction. Here, $n$ is the qubit count of the input circuit while $m$ is the depth (of the input Clifford circuit).}
    \label{tab:pervious_work}
\end{table}

\subsection{Snaking}

Cups and caps are preparation and projection onto the Bell-state $\ket{\phi^+}=(\ket{00}+\ket{11})/\sqrt{2}$. They enable the ``snaking'' of circuits because of their ``ricochet'' property $(A\otimes I)\ket{\phi^+}=(I\otimes A^T)\ket{\phi^+}$ and its conjugate. 
This means that a gate $A$ may be ricocheted between the qubits involved in a cup or cap changing to $A^T$ in the process.

\begin{figure}[h!]
    \centering
    \includegraphics[width=0.95\linewidth]{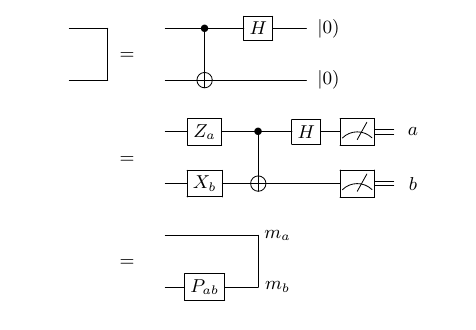}
    \caption{Projection onto the $\ket{\phi^+}$ Bell-state, realized via the Pauli correction $P_{ab}=Z^{m_a}X^{m_b}$.}
    \label{fig:cap_full}
\end{figure}

For the cap (Figure \ref{fig:cap_full}), though true projection is impossible, one may achieve the same circuit via measurement and (classically controlled) Pauli correction. Letting the measurements of the first and second qubit be denoted as $m_a$ and $m_b$, respectively, we may ``cancel out'' undesired measurements (of $\ket{1}$) via the application of Pauli corrections $Z^{a}$ or $X^{b}$ which may be combined into $P_{ab}=Z^{a}X^{b}$ by ricocheting $Z^{m_a}$ from the first qubit to the second qubit.

% \begin{figure}[H]
%     \centering
%     \begin{align*}
% \Qcircuit @C=1em @!R=0.1em{
%     \qw         & \qw & \qw \qwx[1] &                     & & & \ctrl{1}      & \gate{H}    & \qw                 & \hspace{-3em} \stick{|0)} &           \\ 
%     \qw         & \qw & \qw         & \raisebox{2.5em}{=} & & & \targ         & \qw         & \qw                 & \hspace{-3em} \stick{|0)} &           \\ 
%                 &     &             &                     & & & \gate{Z_a}    & \ctrl{1}    & \gate{H}            & \meter                    & \cw & a \\
%                 &     &             & \raisebox{2.5em}{=} & & & \gate{X_b}    & \targ       & \qw                 & \meter                    & \cw & b \\ 
%                 &     &             &                     & & & \qw           & \qw \qwx[1] & \hspace{-3em} m_a   &                           &           \\
%                 &     &             & \raisebox{2.5em}{=} & & & \gate{P_{ab}} & \qw         & \hspace{-3em} m_b   &                           &           \\
%     }
%     \end{align*}
%     \caption{Cap: Projection onto the $\ket{\phi^+}$ Bell-state, realized via the Pauli correction $P_{ab}=Z^{m_a}X^{m_b}$.}
%     \label{fig:cap_full}
% \end{figure}

\subsection{The Zag}
\label{appendix:zag}

\begin{figure}[h!]
    \centering
    \includegraphics[width=0.95\linewidth]{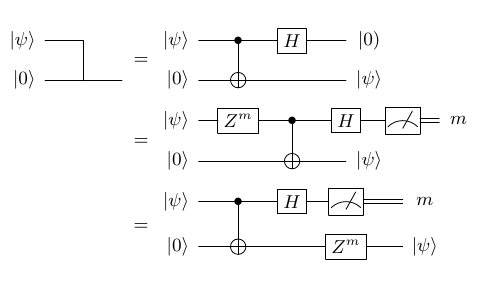}
    \caption{A zag teleports a single qubit state from the first to the second qubit of the diagram.}
    \label{fig:zag}
\end{figure}

A zag (Figure \ref{fig:zag}), also referred to as single-qubit teleportation, is akin to ``half'' of a cap, in which the second qubit is prepared to the $\ket{0}$ state and only the first qubit is projected. Again, though true projection is impossible, an equivalent circuit is achieved via an $X^m$ Pauli-correction, which becomes $Z^m$ when pulled to the front of circuit. Finally, the correction may be teleported to the end of the second qubit's quantum wire.

% \begin{figure}[H]
%     \begin{align*}
%     \label{eq:zag}
%     \Qcircuit @C=1em @R=1em{                   
%     \lstick{\ket{\psi}} & \qw & \qw \qwx[1] &     &     &                    & & & \lstick{\ket{\psi}} & \ctrl{1}   & \gate{H} & \qw        & \hspace{-3.5em} \stick{|0)} &                            &   \\
%     \lstick{\ket{0}}    & \qw & \qw         & \qw & \qw & \raisebox{2em}{=}  & & & \lstick{\ket{0}}    & \targ      & \qw      & \qw        & \hspace{-3.5em} \ket{\psi}  &                            &   \\
%                         &     &             &     &     &                     & & & \lstick{\ket{\psi}} & \gate{Z^m} & \ctrl{1} & \gate{H}   & \meter                      & \cw                        & m \\
%                         &     &             &     &     &  \raisebox{2em}{=}  & & & \lstick{\ket{0}}    & \qw        & \targ    & \qw        & \hspace{-3.5em} \ket{\psi}  &                            &   \\
%                         &     &             &     &     &                     & & & \lstick{\ket{\psi}} & \ctrl{1}   & \gate{H} & \meter     & \cw                         & \hspace{-1.5em} m          &   \\
%                         &     &             &     &     &  \raisebox{2em}{=}  & & & \lstick{\ket{0}}    & \targ      & \qw      & \gate{Z^m} & \qw                         & \hspace{-1.5em} \ket{\psi} &   \\
%     }
%     \end{align*}
%     \caption{A zag teleports a single qubit state from the first to the second qubit of the diagram.}
%     \label{fig:zag}
% \end{figure}

\begin{figure}[h!]
    \centering
    \includegraphics[width=0.95\linewidth]{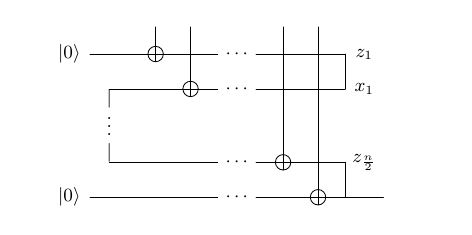}
    \caption{The snaking of the ancilla qubits of Figure \ref{fig:pauli_phasor} (RHS) for even weight Pauli-phasors. The bottom cap is replaced with a zag, as defined in Figure \ref{fig:zag}. The formulae for $\theta_x$ and $s_j$ stay the same from Figure \ref{fig:pauli_phasor}.}
    \label{fig:even_ancilla}
\end{figure}

Finally, the RHS of Figure \ref{fig:even_ancilla} shows how the zag is used to snake the ancilla of an even weight Pauli-phasor. Note that, without this technique, even weight Pauli-phasors would require $n+1$ ancillas for snaking (LHS of Figure \ref{fig:even_ancilla}), with the last ancilla not having a CX target applied to it and simply being used to ensure that the snake ends ``moving forward''.

\subsection{Partial Ricocheting}
\label{sec:partial_ricochet}

The targets of the CNOTs in Figure \ref{fig:pauli_phasor} are able to be ricocheted because CNOT is equal to its own partial transpose (on the $\mathcal{H}_1$ subspace of gates Hilbert space $\mathcal{H}_0\otimes\mathcal{H}_1$):
\begin{align}
\text{CNOT}^{T_1} 
&=
\ket{0}\bra{0}\otimes I^T + \ket{1}\bra{1} \otimes X^T 
\nonumber
\\
&= \ket{0}\bra{0}\otimes I + \ket{1}\bra{1} \otimes X
\nonumber
\\
&=\text{CNOT}.
\end{align}

\begin{figure*}
    \centering
    \includegraphics[width=\linewidth]{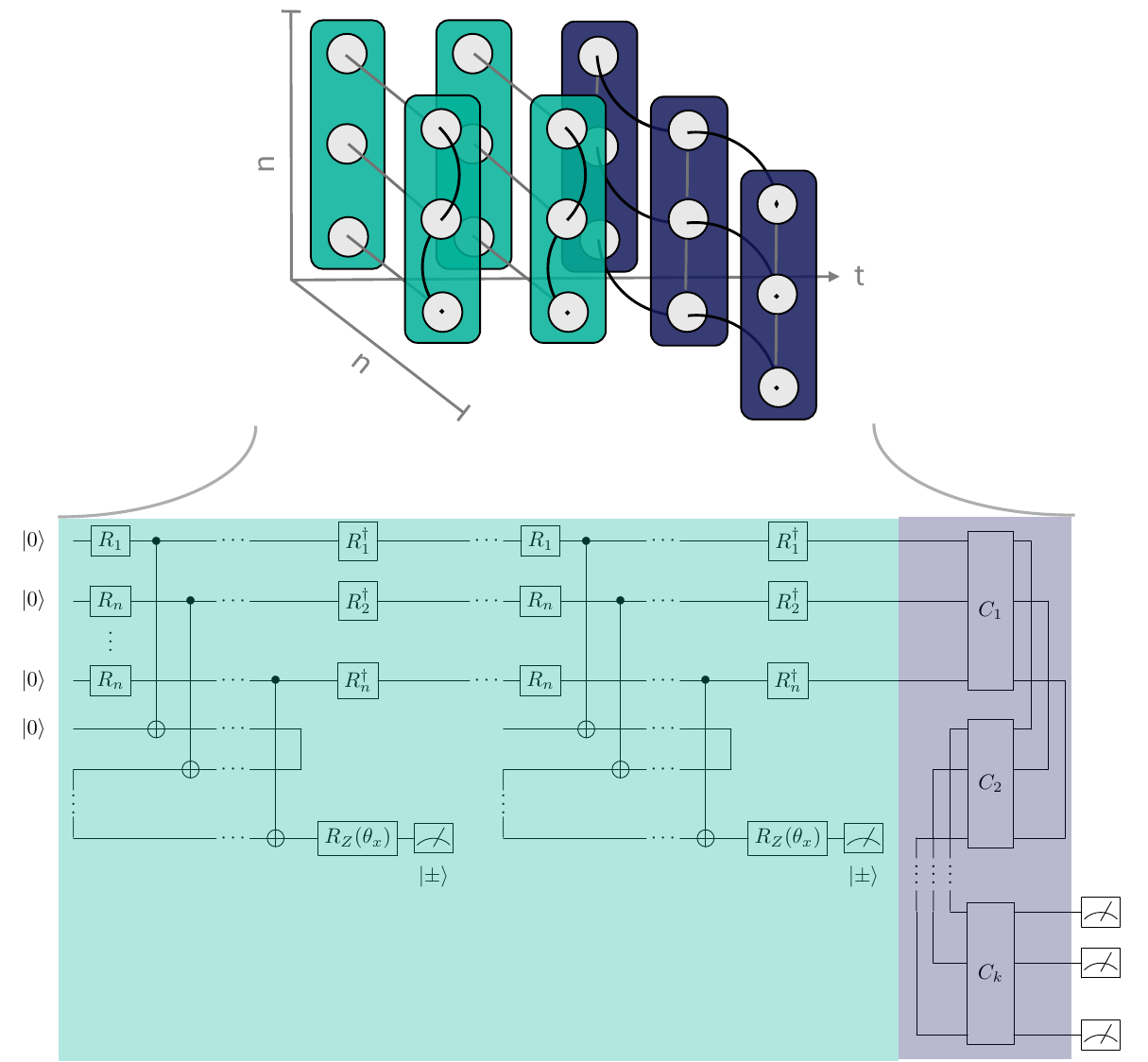}
    \caption{
    (Top) ``Zoomed in'' reduce diagram (same as Figure \ref{fig:reduce}).
    (Bottom) Circuit structure after QDC compilation.
    Both depictions consist of a series of reduced Pauli-phasors (turquoise) followed by the reduced Clifford section (blue).
    }
    \label{fig:qdc_full}
\end{figure*}

\subsection{QDC Circuit Structure}

Figure \ref{fig:qdc_full} above depicts the circuit structure at the end of QDC compilation. 
The top diagram depicts time on the horizontal axis and qubit grid size on the other two axes. 
The rectangles represent columns of the grid.
The turquoise and blue colors represent reduced non-Clifford Pauli-phasors $P''_i$ and the reduced Clifford section $C''$, respectively. 
The gray circles represent qubits of an $n$ x $n$ grid. Note that for the Pauli-phasor time steps, the last column of qubits has been omitted for ease of visualization. In this case, $n=3$ and there are two Pauli-phasors but these quantities generalize to a larger grid and more turquoise time steps, respectively.
The straight gray lines represent the CX gates of the original Pauli-phasor and Clifford sections. 
The curved black lines represent ``snaking'' cups and caps (Figure \ref{fig:cup_and_cap}).

The bottom diagram gives the explicit circuit structure, shaded with colors that have the same meaning as in the top diagram. Note that, as compared to Figure \ref{fig:pauli_phasor}, all data qubit corrections $Z^{s_i}$ have been pushed off the end of the circuit and into post-processing. Similarly, the reset gates, which would normally be applied to the ancilla qubits after each Pauli-phasor, have been replaced with $X$ corrections which have similarly been pushed into post-processing. The cups and caps of the bottom diagram directly correspond to the curved black lines of the top diagram. Additionally, note that, unlike Figure \ref{fig:reduce_cliff}, the reduced Clifford section here has all $n$ qubit wires, showing the grid connectivity of the circuit. Finally, note that only the final column of qubits  needs to be measured at the end of the circuit. The measurements of the first $n-1$ columns inform the post-processing of the the final column's measurements.

\bibliography{bib.bib}

@article{ibm_gate_speed,
  title={Superconducting quantum computers: who is leading the future?},
  author={Muhammad AbuGhanem},
  journal={EPJ Quantum Technology},
  volume={12},
  pages={102},
  year={2025},
  publisher={SpringerOpen},
  doi={10.1140/epjqt/s40507-025-00405-7},
  url={https://epjquantumtechnology.springeropen.com/articles/10.1140/epjqt/s40507-025-00405-7}
}

@misc{cafqa,
      title={CAFQA: A classical simulation bootstrap for variational quantum algorithms}, 
      author={Gokul Subramanian Ravi and Pranav Gokhale and Yi Ding and William M. Kirby and Kaitlin N. Smith and Jonathan M. Baker and Peter J. Love and Henry Hoffmann and Kenneth R. Brown and Frederic T. Chong},
      year={2023},
      eprint={2202.12924},
      archivePrefix={arXiv},
      primaryClass={quant-ph},
      url={https://arxiv.org/abs/2202.12924}, 
}

@article{cliff_synth,
   title={CNOT circuits need little help to implement arbitrary Hadamard-free Clifford transformations they generate},
   volume={9},
   ISSN={2056-6387},
   url={http://dx.doi.org/10.1038/s41534-023-00760-2},
   DOI={10.1038/s41534-023-00760-2},
   number={1},
   journal={npj Quantum Information},
   publisher={Springer Science and Business Media LLC},
   author={Maslov, Dmitri and Yang, Willers},
   year={2023},
   month=sep}

@article{neutral_atom_qubit_record,
  title   = {A tweezer array with 6,100 highly coherent atomic qubits},
  author  = {Manetsch, Hannah J. and Nomura, Gyohei and Bataille, Elie and Leung, Kon H. and Lv, Xudong and Endres, Manuel},
  journal = {Nature},
  volume  = {647},
  pages   = {60--67},
  year    = {2025},
  doi     = {10.1038/s41586-025-09641-4},
  url     = {https://www.nature.com/articles/s41586-025-09641-4},
}

@article{neutral_atom_gate_speed,
  title   = {A universal neutral‑atom quantum computer with individual optical addressing and nondestructive readout},
  author  = {Radnaev, A. G. and Chung, W. C. and Cole, D. C. and Mason, D. and Ballance, T. G. and Bedalov, M. J. and Belknap, D. A. and Berman, M. R. and Blakely, I. L. and Bloomfield, I. L. and Buttler, P. D. and Campbell, C. and Chopinaud, A. and Copenhaver, E. and Dawes, M. K. and Eubanks, S. Y. and Friss, A. J. and Garcia, D. M. and Gilbert, J. and Gillette, M. and Goiporia, P. and Gokhale, P. and Goldwin, J. and Goodwin, D. and Graham, T. M. and Guttormsson, C. J. and Hickman, G. T. and Hurtley, L. and Iliev, M. and Jones, E. B. and Jones, R. A. and Kuper, K. W. and Lewis, T. B. and Lichtman, M. T. and Majdeteimouri, F. and Mason, J. J. and McMaster, J. K. and Miles, J. A. and Mitchell, P. T. and Murphree, J. D. and Neff‑Mallon, N. A. and Oh, T. and Omole, V. and Parlo Simon, C. and Pederson, N. and Perlin, M. A. and Reiter, A. and Rines, R. and Romlow, P. and Scott, A. M. and Stiefvater, D. and Tanner, J. R. and Tucker, A. K. and Vinogradov, I. V. and Warter, M. L. and Yeo, M. and Saffman, M. and Noel, T. W.},
  journal = {PRX Quantum},
  volume  = {6},
  number  = {3},
  pages   = {030334},
  year    = {2025},
  doi     = {10.1103/66s8‑jj18},
  url     = {https://journals.aps.org/prxquantum/abstract/10.1103/66s8‑jj18}
}

@article{mbqc,
  title   = {A one‑way quantum computer},
  author  = {Raussendorf, Robert and Briegel, Hans J.},
  journal = {Physical Review Letters},
  volume  = {86},
  number  = {22},
  pages   = {5188--5191},
  year    = {2001},
  doi     = {10.1103/PhysRevLett.86.5188},
  url     = {https://doi.org/10.1103/PhysRevLett.86.5188}
}

@article{gate_teleportation,
  title   = {Demonstrating the viability of universal quantum computation using teleportation and single‑qubit operations},
  author  = {Gottesman, Daniel and Chuang, Isaac L.},
  journal = {Nature},
  volume  = {402},
  pages   = {390--393},
  year    = {1999},
  doi     = {10.1038/46503},
  url     = {https://doi.org/10.1038/46503}
}

@article{mcclean2018barren,
  title        = {Barren plateaus in quantum neural network training landscapes},
  author       = {McClean, Jarrod R. and Boixo, Sergio and Smelyanskiy, Vadim N. and Babbush, Ryan and Neven, Hartmut},
  journal      = {Nature Communications},
  volume       = {9},
  number       = {1},
  pages        = {4812},
  year         = {2018},
  publisher    = {Nature Publishing Group},
  doi          = {10.1038/s41467-018-07090-4},
  eprint       = {1803.11173},
  archivePrefix= {arXiv},
  primaryClass = {quant-ph}
}

@article{holmes2022expressibility,
  title        = {Connecting Ansatz Expressibility to Gradient Magnitudes and Barren Plateaus},
  author       = {Holmes, Zoë and Sharma, Kunal and Cerezo, Marco and Coles, Patrick J.},
  journal      = {PRX Quantum},
  volume       = {3},
  number       = {1},
  pages        = {010313},
  year         = {2022},
  publisher    = {American Physical Society},
  doi          = {10.1103/PRXQuantum.3.010313},
  eprint       = {2101.02138},
  archivePrefix= {arXiv},
  primaryClass = {quant-ph}
}

@phdthesis{gottesman1997stabilizer,
  author       = {Daniel Gottesman},
  title        = {Stabilizer Codes and Quantum Error Correction},
  school       = {California Institute of Technology},
  year         = {1997},
  eprint       = {quant-ph/9705052},
  archivePrefix= {arXiv},
  primaryClass = {quant-ph}
}

@article{qaoa,
  title={A Quantum Approximate Optimization Algorithm},
  author={Farhi, Edward and Goldstone, Jeffrey and Gutmann, Sam},
  journal={arXiv preprint arXiv:1411.4028},
  year={2014}
}

@article{vqe,
  title={A variational eigenvalue solver on a photonic quantum processor},
  author={Peruzzo, Alberto and McClean, Jarrod and Shadbolt, Peter and Yung, Man-Hong and Zhou, Xiao-Qi and Love, Peter J and Aspuru-Guzik, Al{\'a}n and O’Brien, Jeremy L},
  journal={Nature Communications},
  volume={5},
  pages={4213},
  year={2014},
  publisher={Nature Publishing Group}
}

@article{qml,
  title={Quantum machine learning},
  author={Biamonte, Jacob and Wittek, Peter and Pancotti, Nicola and Rebentrost, Patrick and Wiebe, Nathan and Lloyd, Seth},
  journal={Nature},
  volume={549},
  number={7671},
  pages={195--202},
  year={2017},
  publisher={Nature Publishing Group}
}

@article{qrbm,
  title   = {Expressive equivalence of classical and quantum restricted Boltzmann machines},
  author  = {Demidik, Maria and T{\"u}ys{\"u}z, Cenk and Piatkowski, Nico and Grossi, Michele and Jansen, Karl},
  journal = {Communications Physics},
  volume  = {8},
  pages   = {413},
  year    = {2025},
  doi     = {10.1038/s42005-025-02353-1},
  url     = {https://doi.org/10.1038/s42005-025-02353-1}
}

@article{heisenberg,
  title   = {Zur Theorie des Ferromagnetismus},
  author  = {Heisenberg, Werner},
  journal = {Zeitschrift f{\"u}r Physik},
  volume  = {49},
  number  = {9-10},
  pages   = {619--636},
  year    = {1928},
  doi     = {10.1007/BF01328601},
  url     = {https://doi.org/10.1007/BF01328601}
}

@article{buhrman2024,
   title={State preparation by shallow circuits using feed forward},
   volume={8},
   ISSN={2521-327X},
   url={http://dx.doi.org/10.22331/q-2024-12-09-1552},
   DOI={10.22331/q-2024-12-09-1552},
   journal={Quantum},
   publisher={Verein zur Forderung des Open Access Publizierens in den Quantenwissenschaften},
   author={Buhrman, Harry and Folkertsma, Marten and Loff, Bruno and Neumann, Niels M. P.},
   year={2024},
   month=dec, pages={1552}}

@article{yang2023,
  title        = {Harnessing the Power of Long-Range Entanglement for Clifford Circuit Synthesis},
  author       = {Yang, Willers and Rall, Patrick},
  journal      = {arXiv preprint arXiv:2302.06537},
  year         = {2023},
  archivePrefix= {arXiv},
  eprint       = {2302.06537},
  primaryClass = {quant-ph},
  url          = {https://arxiv.org/abs/2302.06537}
}

@misc{quclear,
      title={QuCLEAR: Clifford Extraction and Absorption for Quantum Circuit Optimization}, 
      author={Ji Liu and Alvin Gonzales and Benchen Huang and Zain Hamid Saleem and Paul Hovland},
      year={2025},
      eprint={2408.13316},
      archivePrefix={arXiv},
      primaryClass={quant-ph},
      url={https://arxiv.org/abs/2408.13316}, 
}

@misc{Qiskit,
author       = {H{'e}ctor Abraham and AduOffei and Ismail Yunus Akhalwaya and
Gadi Aleksandrowicz and Thomas Alexander and et al.},
title        = {{Qiskit: An Open-source Framework for Quantum Computing}},
year         = {2019},
howpublished = {\url{https://qiskit.org}},
doi          = {10.5281/zenodo.2562110}
}

\end{document}